\documentclass[final,11pt,number,sort&compress]{elsarticle}
\usepackage{graphicx}
\usepackage{amsmath}
\journal{Chemical Physics Letters}

\begin{document}
\begin{frontmatter}

\title{Choosing a density functional for static molecular polarizabilities}
\author{Taozhe Wu},
\author{Yulia~N.~Kalugina\corref{perm}},
\ead{kalugina@phys.tsu.ru}
\author{Ajit~J.~Thakkar\corref{cor}}
\ead{ajit@unb.ca}
\address{Department of Chemistry, University of New Brunswick,\\
Fredericton, New Brunswick E3B 5A3, Canada}
\cortext[perm]{Permanent address: Tomsk State University, 36 Lenin Av., Tomsk 634050, Russia}
\cortext[cor]{Corresponding author}

\begin{abstract}
Coupled-cluster calculations of static electronic dipole polarizabilities for 145 organic molecules
are performed to create a reference data set. The molecules are composed from carbon, hydrogen,
nitrogen, oxygen, fluorine, sulfur, chlorine, and bromine atoms. They range in size from triatomics
to 14 atoms. The Hartree-Fock and 2nd-order M{\o}ller-Plesset methods and 34 density functionals,
including local functionals, global hybrid functionals, and range-separated functionals of the
long-range-corrected and screened-exchange varieties, are tested against this data set. On the basis
of the test results, detailed recommendations are made for selecting density functionals for polarizability
computations on relatively small organic molecules.
\end{abstract}

\end{frontmatter}

\section{Introduction}\label{sec:intro}

Orbital-dependent density functional theory (DFT)~\cite{Kummel08} is the method of choice in many
studies of electronic structure, particularly those involving systems too large to handle with accurate
wave function methods such as the coupled-cluster approach~\cite{Bartlett07} and large basis sets.
However, the exact exchange-correlation functional $E_{\text{xc}}$ remains, and likely will remain,
elusive. Hence, many different approximate $E_{\text{xc}}$ are continually being invented. In turn, this
creates the need for assessment and benchmark studies; see, for example, Peverati and Truhlar's recent
comprehensive assessment of the accuracy of 77 density functional methods~\cite{Peverati14}.

Assessments of the accuracy of density functionals for the calculation of polarizabilities and
hyperpolarizabilities have appeared periodically; see, for example,
Refs.~\cite{Adamo99b,Cohen99,Wilson01,Jacquemin07,Limacher09,Laczkowska13}. The purpose of this
Letter is to report how well a comprehensive range of functionals, including recently proposed ones,
predict the static mean electronic dipole polarizabilities ($\alpha$) of molecules of interest in
bioorganic and pharmaceutical chemistry. Atomic units are used throughout. The SI value of one
atomic unit of the dipole polarizability $\alpha$ is $1.648\,777
\times{10^{-41}}\,\mathrm{C^{2}\,m^{2}\,J^{-1}}$ on the basis of the 2010 CODATA recommendations for
the fundamental physical constants~\cite{CODATA10}.

\section{Methods and calculations}\label{sec:meth}

A large set of reference polarizability data is required to make a robust assessment. The first
thought is to use good quality gas-phase experimental data. A recent survey~\cite{Hohm13} shows that
measured gas-phase dipole polarizabilities are known for 166 molecules, not counting isotopologues
as distinct. Although data with a precision of 0.1\% is available for a few small molecules,
Hohm~\cite{Hohm13} explains that the experimental error is much greater in many cases. Moreover, he
warns~\cite{Hohm13} that ``there are also molecules like HI, CH$_{2}$Cl$_{2}$ or CH$_{2}$Br$_{2}$
for which the available data do not even overlap within their error bars''. Further, the need to
remove vibrational contributions from the experimental values adds extra uncertainties.

In view of the apparent difficulty in constructing a large set of reliable measured static
electronic dipole polarizabilities, we decided to use a reference data set computed using a
high-level ab initio method---the CCSD(T) coupled-cluster approach using single and double
substitutions with a non-iterative correction for triple
substitutions~\cite{Raghavachari89,Bartlett07}. The CCSD(T) method is less accurate for molecules
that have strong multi-configuration character. However, it should be sufficiently accurate for
creating a test set for density functional methods.

An augmented, correlation-consistent, polarized valence-triple-zeta (aug-cc-pVTZ) basis
set~\cite{Kendall92,Woon93a} was used for all calculations reported in this work. A quadruple-zeta
or larger basis set would yield results closer to the basis set limit. However, the size of the
basis set constrains the size of the molecules that can be included in the test set. The choice of
the aug-cc-pVTZ basis is a pragmatic compromise that allows us to include molecules of moderate size
as described in the next paragraph. Moreover, atoms and diatomics are excluded from the test set to
bypass extreme basis set effects seen in small systems; see, for example, Ref.~\cite{Thakkar05na}.
This work which consistently uses the aug-cc-pVTZ basis set for all calculations should be viewed as
an assessment in a model chemistry in Pople's sense~\cite{Pople73}.

Next, the reference set of molecules must be chosen. In a recent study of additive models for
polarizabilities~\cite{Blair14add}, a training set of 298 molecules (T298) was selected carefully
from the TABS database~\cite{Blair14TABS}. The latter contains computed equilibrium structures of
1641 organic molecules with at least 25 molecules in each of 24 functional categories.
Unfortunately, about half of the molecules in T298 are too large even for CCSD(T)/aug-cc-pVTZ
calculations with our computational resources. Thus, we chose 145 of the smallest molecules from
T298. This subset is referred to as T145 in this work. As in the parent TABS
database~\cite{Blair14TABS} and T298 training set~\cite{Blair14add}, the molecules in T145 contain
at least one carbon atom and possibly one or more of the H, N, O, F, S, Cl, and Br atoms. The
molecules in T145 contain an average of 9.1 atoms with an average of 5.5 being non-hydrogen atoms.
The smallest molecule in T145 contains three atoms (FCN) and the largest contains 14 atoms
(1,4-dithiane). The molecules in T145 contain an average of 51 electrons ranging from 18 electrons
in CH$_{3}$F to 120 electrons in CBr$_{3}$CH$_{3}$. Ball-and-stick diagrams and TABS identification
(ID) numbers for the molecules in T145 are included in the supplementary material.

Since hybrid density functionals generally exceed the performance of local ones for
molecules~\cite{Kummel08,Peverati14}, only a few representative local functionals were selected. We
consider the local spin-density approximation (LSDA) with the VWN5 fit to the electron gas
correlation energy~\cite{Vosko80}, and two generalized (G) gradient approximations (GA):
PBE~\cite{Perdew96} and HCTH~\cite{Boese01}. We also examine a non-separable gradient approximation
(NGA)~\cite{Peverati14}, the recent N12~\cite{Peverati12a} functional. As representative functionals
that use the kinetic-energy density, we examine the $\tau$HCTH~\cite{Boese02},
revTPSS~\cite{Perdew09}, and M11-L~\cite{Peverati12b} meta-GGAs, and the MN12-L~\cite{Peverati12d}
meta-NGA.

Global hybrid (GH) functionals mix a constant fraction of HF exchange with gradient approximations
to the exchange energy. These can be subdivided into those that mix a GGA and those that mix a
meta-GGA with HF exchange. Following Peverati and Truhlar~\cite{Peverati14}, we refer to these kinds
as GH-GGA and GH-mGGA, respectively. As representative of the GH-GGA variety, we examine the
archetypal B3PW91~\cite{Becke93a}, the popular B3LYP~\cite{Stephens94}, PBE0 (also known as
PBE1PBE)~\cite{Ernzerhof99,Adamo99}, mPW1PW~\cite{Adamo98}, B97-2~\cite{Wilson01}, and
SOGGA11-X~\cite{Peverati11} functionals. As representative of the GH-mGGA variety, we examine the
$\tau$HCTHhyb~\cite{Boese02}, TPSSh~\cite{Staroverov03}, BMK~\cite{Boese04}, M06-HF~\cite{Zhao06b},
M06~\cite{Zhao08}, and M06-2X~\cite{Zhao08} functionals.

Range-separated hybrid (RSH) functionals mix different forms of exchange approximations in amounts
that vary with the interelectronic separation~\cite{Leininger97,Iikura01}. Such functionals can be
quite effective for polarizabilities~\cite{Jacquemin07,Limacher09,Laczkowska13}. The ones we examine
include CAM-B3LYP~\cite{Yanai04}, LC-$\omega$PBE~\cite{Vydrov06}, HISS~\cite{Henderson07},
$\omega$B97~\cite{Chai08a}, $\omega$B97X~\cite{Chai08a}, $\omega$B97X-D~\cite{Chai08},
HSE06~\cite{Henderson09}, M11~\cite{Peverati11b}, N12-SX~\cite{Peverati12e}, and
MN12-SX~\cite{Peverati12e}. Of these, only the M11 and MN12-SX functionals use the kinetic-energy
density. The long-range correction (LC) scheme of Iikura et al.~\cite{Iikura01} can be used to
produce an RSH functional from any local functional. We used this procedure to test four less-known
RSH functionals, LC-HCTH, LC-$\tau$HCTH, LC-revTPSS, and LC-N12, generated from what turn out to be
the four best local functionals.

Most DFT computations were carried out with an `ultra-fine', pruned (99,590) Lebedev numerical
integration grid of 99 radial and 590 angular points. However, a larger (150,974) [(225,974) for Br]
`super-fine' grid,  was used in cases where the density functional has a dependence on the kinetic
energy density.

Hartree-Fock (HF) and 2nd-order M{\o}ller-Plesset (MP2) polarizabilities were also computed because
they are standard reference points for wave function methods in quantum chemistry. All calculations
were carried out using Gaussian 09~\cite{g09s} and the MOLPRO 2010.1 packages~\cite{Molpro10}.
Tables of the calculated polarizabilities can be found in the supplementary material.

\section{Assessment of functionals}\label{sec:ass}

An assessment and ranking of the density functionals requires some criteria and, if possible,
associated numerical measures. The absolute percent deviations of the DFT polarizabilities from
their CCSD(T) counterparts can be used in many ways to construct such measures. Five numerical
indices constructed from these deviations are the median (MED), average (AVE), root mean square
(RMS), 90th percentile ($P_{90}$), and maximum (MAX) as listed in Table~1. Spurious numerical
artifacts may arise with percent deviations if the quantities themselves are small. However, there
is no cause for concern in this work because the polarizabilities of the 145 molecules considered
range in magnitude from 16.8 au to 116 au.

\begin{table}[tbp]
\centering
\caption{Median (MED), average (AVE), root mean square (RMS), 90th percentile ($P_{90}$), and
maximum (MAX) absolute percent deviations of the polarizabilities predicted by the various methods
with respect to the CCSD(T) polarizabilities. $B$ is the bias parameter and $\Delta= P_{90} +
\vert{B}\vert$. All quantities are unitless.}
\begin{tabular}{lrrrrrrr}
\hline
Method & $B$ & MED & AVE & RMS & $P_{90}$ & MAX & $\Delta$ \\
\hline
      MP2         & $ 0.63$ &   0.47  &   0.53  &   0.64  &   1.02  &   1.83  &    1.6 \\
      M11         & $ 0.08$ &   1.04  &   1.33  &   1.68  &   2.83  &   4.98  &    2.9 \\
   M06-2X         & $-0.01$ &   0.93  &   1.24  &   1.69  &   2.85  &   5.68  &    2.9 \\
     $\omega$B97  & $-0.12$ &   1.15  &   1.45  &   1.77  &   2.96  &   4.60  &    3.1 \\
 LC-$\tau$HCTH    & $-0.02$ &   1.14  &   1.38  &   1.82  &   3.16  &   7.03  &    3.2 \\
   M06-HF         & $-0.42$ &   1.64  &   1.75  &   2.10  &   2.95  &   5.98  &    3.4 \\
    $\omega$B97X  & $ 0.19$ &   1.07  &   1.41  &   1.81  &   3.35  &   5.20  &    3.5 \\
  LC-HCTH         & $-0.57$ &   1.58  &   1.76  &   2.13  &   3.52  &   5.20  &    4.1 \\
     HISS         & $-0.56$ &   1.73  &   1.87  &   2.27  &   3.63  &   6.75  &    4.2 \\
CAM-B3LYP         & $ 0.49$ &   1.11  &   1.51  &   2.07  &   3.80  &   6.24  &    4.3 \\
SOGGA11-X         & $ 0.30$ &   0.94  &   1.50  &   2.14  &   4.07  &   7.15  &    4.4 \\
   $\omega$B97XD  & $ 0.49$ &   1.25  &   1.59  &   2.14  &   4.07  &   6.27  &    4.6 \\
  LC-$\omega$PBE  & $-0.68$ &   2.09  &   2.01  &   2.38  &   3.89  &   5.51  &    4.6 \\
      BMK         & $ 0.26$ &   1.10  &   1.62  &   2.25  &   4.40  &   7.97  &    4.7 \\
   N12-SX         & $ 0.14$ &   0.93  &   1.58  &   2.35  &   4.52  &   8.39  &    4.7 \\
 LC-revTPSS       & $-0.81$ &   2.64  &   2.50  &   2.91  &   4.50  &   6.41  &    5.3 \\
   mPW1PW         & $ 0.61$ &   1.03  &   1.75  &   2.46  &   4.84  &   8.31  &    5.4 \\
     PBE0         & $ 0.75$ &   1.28  &   1.96  &   2.62  &   5.01  &   8.51  &    5.8 \\
    B97-2         & $ 0.71$ &   1.27  &   1.95  &   2.67  &   5.13  &   8.90  &    5.8 \\
    HSE06         & $ 0.77$ &   1.49  &   2.09  &   2.75  &   5.17  &   8.90  &    5.9 \\
   B3PW91         & $ 0.77$ &   1.51  &   2.14  &   2.81  &   5.34  &   9.00  &    6.1 \\
      M06         & $ 0.94$ &   2.62  &   2.89  &   3.29  &   5.13  &   7.70  &    6.1 \\
    TPSSh         & $ 0.93$ &   2.48  &   3.01  &   3.61  &   5.61  &  10.74  &    6.5 \\
 $\tau$HCTHhyb    & $ 0.93$ &   2.42  &   2.97  &   3.60  &   5.88  &  10.55  &    6.8 \\
   LC-N12         & $-0.97$ &   4.00  &   3.80  &   4.16  &   5.98  &   8.08  &    6.9 \\
    B3LYP         & $ 0.94$ &   2.57  &   3.10  &   3.62  &   6.25  &  10.20  &    7.2 \\
  revTPSS         & $ 0.97$ &   4.13  &   4.55  &   5.03  &   7.14  &  13.47  &    8.1 \\
       HF         & $-0.82$ &   3.89  &   4.24  &   4.79  &   7.41  &  10.52  &    8.2 \\
      N12         & $ 0.94$ &   3.51  &   3.97  &   4.63  &   7.41  &  13.65  &    8.4 \\
    $\tau$HCTH    & $ 0.96$ &   4.18  &   4.56  &   5.08  &   7.45  &  13.80  &    8.4 \\
     HCTH         & $ 0.97$ &   4.64  &   4.89  &   5.35  &   7.44  &  13.82  &    8.4 \\
  MN12-SX         & $ 0.97$ &   4.22  &   4.82  &   5.30  &   7.97  &  12.76  &    8.9 \\
      PBE         & $ 0.97$ &   5.79  &   6.08  &   6.45  &   8.50  &  15.13  &    9.5 \\
   MN12-L         & $ 0.96$ &   4.27  &   4.90  &   5.60  &   8.82  &  14.37  &    9.8 \\
     LSDA         & $ 0.97$ &   6.24  &   6.55  &   6.90  &   8.95  &  15.38  &    9.9 \\
    M11-L         & $ 0.97$ &   7.64  &   8.32  &   8.78  &  12.89  &  18.07  &   13.9 \\
\hline
\end{tabular}
\end{table}

Information about the signs of the errors can be quantified by working with a bias measure
$B=f_{+}-f_{-}$ in which $f_{+}$ and $f_{-}$ are the fractions of errors that are positive and
negative respectively. The bias ranges from $-1$ to $+1$. A bias $B=-1$ indicates that the
deviations are all negative, $B=0$ indicates no bias since positive and negative errors occur in
equal numbers, and $B=1$ indicates that the deviations are all positive. Different rankings are
obtained from the five different percent deviation measures. However, a unique measure needs to be
chosen to create rankings. A density functional of general utility should be accurate most of the
time. Thus, as low a value as possible of $P_{90}$, the upper bound to the absolute percent
deviation 90\% of the time, is desirable. Moreover, as small a bias as possible is desirable as
well. Hence, we chose to rank the functionals by the smallness of an error measure defined as
$\Delta=P_{90}+\vert{B}\vert$. This choice of $\Delta$ is subjective but so are other choices. The
bias $B$ and $\Delta$ are also listed in Table 1 for the HF, MP2, and 34 density functional methods.
Table 1 is ordered with respect to $\Delta$. The values of $\Delta$ have been rounded to one decimal
because we think that differences smaller than 0.1 in $\Delta$ are not meaningful for ranking
functionals.

By and large, the expected increase in accuracy of functionals as one moves up the rungs of the
`Jacob's ladder' hierarchy~\cite{Perdew01} is reflected in Table 1. The eight local functionals
examined all have values of $\Delta>8$. There is a smooth decrease of $\Delta$ as one progresses
from the LSDA through the PBE GGA and the revTPSS meta-GGA to the global hybrid PBE0, all of which
are in the `reductionist' category of functionals~\cite{Perdew01}. All the local functionals have
large positive bias values of $B\ge{0.94}$ indicating a strong tendency to overestimate the CCSD(T)
polarizability. In addition to looking at overall numerical indicators, it is instructive to examine
the error distributions visually. Figure~1 shows the signed polarizability errors
$\delta\alpha=\alpha(\mathrm{model})-\alpha(\mathrm{CCSD(T)})$ for the LSDA, revTPSS, and PBE0
functionals; the PBE GGA is omitted from Figure~1 because its errors are quite similar to those of
the revTPSS meta-GGA and would add too much clutter to Figure~1. The large positive bias of the
local functionals is seen clearly in Figure~1. The global hybrid PBE0 has a visibly lesser tendency
($B=0.75$) to overestimate polarizabilities. The $\pm{3}$\% envelope is shown in Figure~1 and all
subsequent figures to give a sense of the distribution of the percent deviations and to facilitate
comparing figures with different vertical scales. The most accurate local functional, the meta-GGA
revTPSS ($\Delta=8.1$), is marginally more accurate than the HF method ($\Delta=8.2$).

\begin{figure}[tbp]
\centering
\includegraphics[scale=0.95]{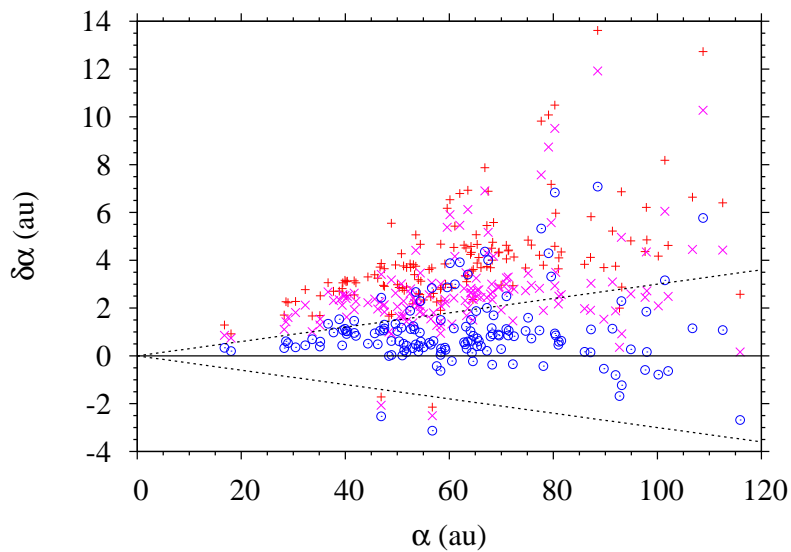}
\caption{Polarizability errors $\delta\alpha$ of the LSDA (+), revTPSS (x), and PBE0 ($\circ$)
density functionals with respect to the CCSD(T) polarizability $\alpha$. The dotted lines correspond to
$\pm{3}$\% of the CCSD(T) $\alpha$. \label{fig:mgga}}
\end{figure}

\begin{figure}[tbp]
\centering
\includegraphics[scale=0.95]{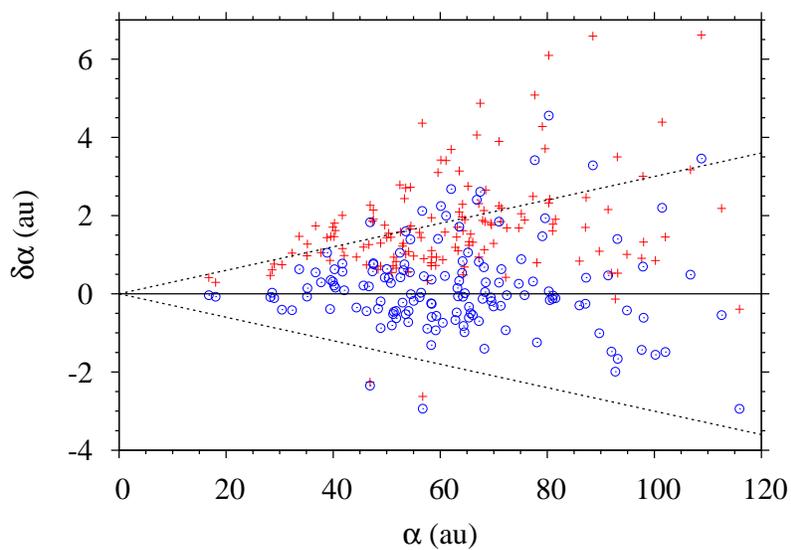}
\caption{Polarizability errors $\delta\alpha$ of the M06 (+) and M06-2X ($\circ$) density
functionals with respect to the CCSD(T) polarizability $\alpha$. The dotted lines correspond to $\pm{3}$\% of
the CCSD(T) $\alpha$. \label{fig:hx}}
\end{figure}

Most of the 12 global hybrid functionals studied in this work predict moderately accurate
polarizabilities. The values of $\Delta$ decrease from 7.2 to 5.4 in the order B3LYP, $\tau$HCTHhyb,
TPSSh, M06, B3PW91, B97-2, PBE0, and mPW1PW for the eight functionals which mix in low to moderate
amounts of HF exchange ($X$) ranging between 10\% and 27\%. The other four global hybrids (BMK,
SOGGA11-X, M06-HF, and M06-2X) are more accurate and have greater percentages ($X\ge{40}\%$) of HF
exchange. Figure~2 compares M06 ($X=27\%$, $\Delta=6.1$) and M06-2X ($X=54\%$, $\Delta=2.9$)
demonstrating clearly how a larger fraction of HF exchange can improve polarizabilities. Of course,
the accuracy of $\alpha$ does not depend solely on $X$; compare M06-2X with M06-HF ($X=100\%$,
$\Delta=3.4$).

A smooth way of reducing self-interaction errors and improving asymptotic behavior is to add
long-range corrections~\cite{Iikura01,Chai08} using range-separation in which the fraction of HF
exchange increases with interelectronic distance $u$. Ten RSH functionals tested in this work do
just that. The amount of HF exchange is increased from 0 at small $u$ to 100\% at large $u$ in six
of them: $\omega$B97, LC-$\omega$PBE, LC-HCTH, LC-$\tau$HCTH, LC-revTPSS, and LC-N12. Three RSH
functionals, $\omega$B97X, $\omega$B97X-D, and M11, include a non-zero amount ($X=16$--$43$\%) of
short-range HF exchange and increase $X$ to 100\% at large $u$. The CAM-B3LYP RSH functional
increases $X$ from 19\% at short range to 65\% at long-range. Table 1 shows that eight of these RSH
functionals have $\Delta\le{4.6}$ and rank among the top dozen for polarizabilities; only LC-revTPSS
and LC-N12 show mediocre performance for $\alpha$. In contrast to the GH functionals, the RSH
functionals of the LC- variety all have a tendency to underestimate the CCSD(T) polarizabilities.
The improvement in $\alpha$ going from the PBE0 GH functional to the LC-$\omega$PBE RSH functional
is illustrated in Figure~3.

\begin{figure}[tbp]
\centering
\includegraphics[scale=0.95]{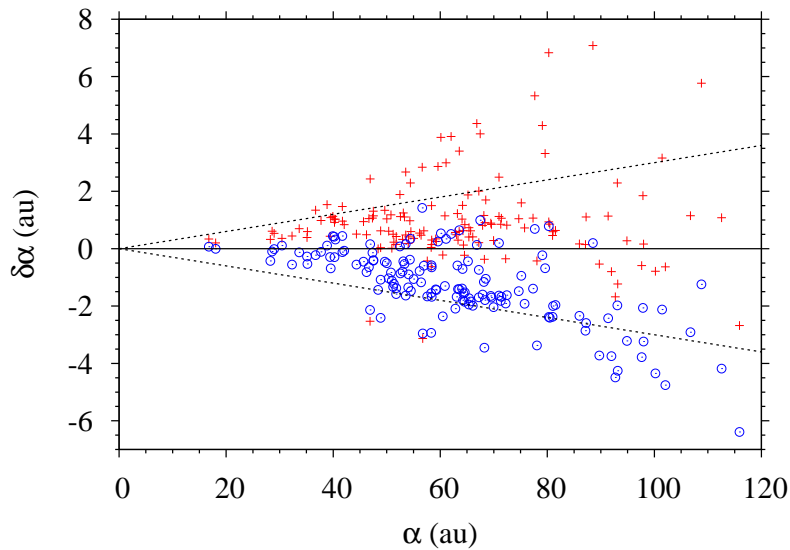}
\caption{Polarizability errors $\delta\alpha$ of the global hybrid PBE0 (+) and long-range-corrected RSH
LC-$\omega$PBE ($\circ$) density functionals with respect to the CCSD(T) polarizability $\alpha$. The
dotted lines correspond to $\pm{3}$\% of the CCSD(T) $\alpha$. \label{fig:lrc}}
\end{figure}

The fine performance of three long-range corrected RSH functionals, M11, $\omega$B97, and
LC-$\tau$-HCTH, is shown in Figure~4. Along with M06-2X, a high-exchange GH functional shown in
Figure~2, these are the very best functionals found in this study for polarizabilities. These four
functionals have $\vert{B}\vert\le{0.12}$ and show no systematic tendency to either overestimate or
underestimate the polarizability. They have the lowest values of $\Delta$, ranging from 2.9 to 3.2,
found for the functionals in Table~1.

\begin{figure}[tbp]
\centering
\includegraphics[scale=0.95]{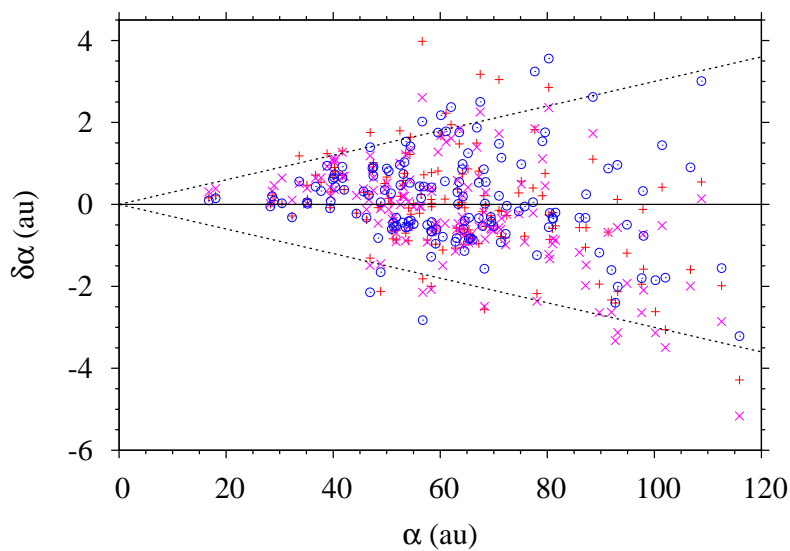}
\caption{Polarizability errors $\delta\alpha$ of the LC-$\tau$HCTH (+), $\omega$B97 (x), and M11
($\circ$) long-range-corrected RSH density functionals with respect to the CCSD(T) polarizability
$\alpha$. The dotted lines correspond to $\pm{3}$\% of the CCSD(T) $\alpha$. \label{fig:best}}
\end{figure}

Screened exchange (SX) RSH functionals in which the amount of HF exchange decays to zero at large
$u$ are more computationally affordable than GH and LC-type RSH functionals for solids. The four
SX-RSH examined (HISS, N12-SX, HSE06, and MN12-SX) have $\Delta$ values ranging from 4.2 to 8.9 and,
as expected, predict $\alpha$ less well than the best long-range corrected RSH functionals. The best
SX-RSH for $\alpha$ is HISS because it maximizes HF exchange at medium- instead of short range.
Interestingly, HISS is of comparable accuracy to CAM-B3LYP which is often
recommended~\cite{Jacquemin07,Limacher09,Laczkowska13} and used for polarizability calculations.
However, screened-exchange functionals like HISS may not perform quite as well for large molecules.
Figure~5 compares the `non-empirical' HISS and HSE06.

\begin{figure}[tbp]
\centering
\includegraphics[scale=0.95]{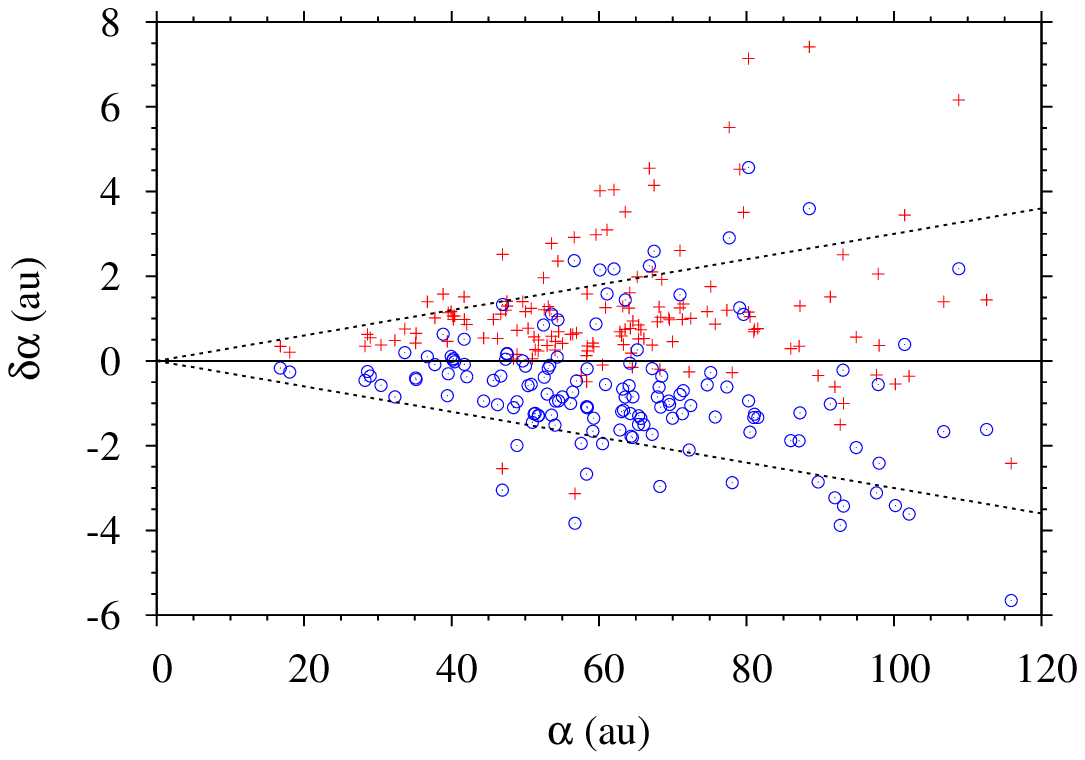}
\caption{Polarizability errors $\delta\alpha$ of the screened-exchange HSE06 (+) and HISS ($\circ$) hybrid
density functionals with respect to the CCSD(T) polarizability $\alpha$. The dotted lines correspond to
$\pm{3}$\% of the CCSD(T) $\alpha$. \label{fig:sx}}
\end{figure}

\begin{figure}[tbp]
\centering
\includegraphics[scale=0.95]{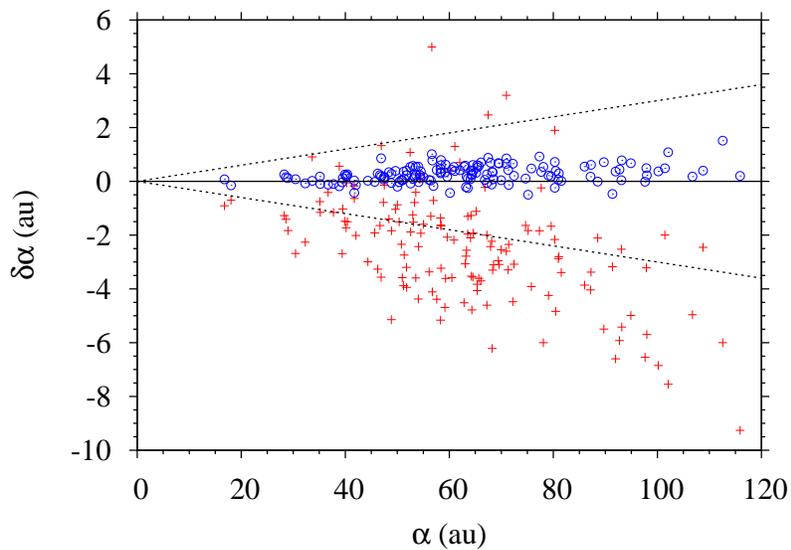}
\caption{Polarizability errors $\delta\alpha$ of the Hartree-Fock (+) and MP2 ($\circ$) methods with
respect to the CCSD(T) polarizability $\alpha$. The dotted lines correspond to $\pm{3}$\% of the CCSD(T)
$\alpha$. \label{fig:wf}}
\end{figure}

Finally, Table~1 shows that the MP2 method is indisputably better for polarizabilities than any of
the density functionals examined. Figure~6 and the bias parameters show that the Hartree-Fock and
MP2 methods have a tendency to under- and over- estimate the polarizability, respectively. The MP2
values scaled by 0.996 give a small improvement over their unscaled counterparts: $B=0.05$, MED = 0.32\%,
$P_{90}=0.79\%$, and $\Delta=0.8$.

\section{Concluding remarks}

Now that an assessment of density functionals has been made, it is appropriate to make
recommendations on the `optimal' choice of density functional(s) for polarizability calculations on
small to medium-size organic molecules. It is imprudent to rely on a single DFT method. We think
that at least three functionals should be used. Two of the DFT methods should be chosen from the top
four performers in Table~1. These four divide neatly into two pairs. One pair consists of M11 and
M06-2X which have the same provenance and the other pair consists of $\omega$B97 and LC-$\tau$HCTH
both of which are related to HCTH and to LC-HCTH. To add greater robustness to the mix, the third
should be one of the two best non-empirical functionals identified in Section~3.

More specifically, we suggest using one of M11 and M06-2X, one of $\omega$B97 and LC-$\tau$HCTH, and
one of HISS and LC-$\omega$PBE. The usual choice in each pair would be the first one named. M11
would be preferred most of the time because it was designed~\cite{Peverati11b,Peverati14} to replace
M06-2X and M06-HF among others, and because the parametrization of M06-2X was limited to molecules
composed solely of non-metal atoms. However, M06-2X can be useful in those situations where M11
suffers from slow convergence or other instabilities~\cite{Peverati12b,Mardirossian13}. The
$\omega$B97 functional is preferred over LC-$\tau$HCTH because the parameters of the former, unlike
the latter, were optimized with the long-range corrections in place. However, the use of the kinetic
energy density in LC-$\tau$HCTH may make it more reliable in some cases. The HISS functional is
preferred over LC-$\omega$PBE because of its better performance for polarizabilities; see Table~1.
However, the retention of HF exchange at long-range may make LC-$\omega$PBE more accurate than HISS
for large molecules. If there are significant differences among the polarizabilities obtained with
the three selected functionals, then MP2 and a sequence of coupled-cluster methods are required if
at all possible with the computational resources at hand.

The above recommendations are being tested by applying them to the molecules for which experimental
gas-phase data is available~\cite{Hohm13}. We expect to report the results in the very near future.

The difficulty in creating a density functional suitable for all properties is attested to by the
observation that none of the functionals mentioned in the recommendations above is among the seven
best general-purpose functionals identified in a recent survey~\cite{Peverati14}.

\section*{Acknowledgments}\label{sec:ack}

The Natural Sciences and Engineering Research Council (NSERC) of Canada supported this work in part. Taozhe Wu
thanks NSERC for an undergraduate summer research award and the University of New Brunswick for a
work-study award. The coupled-cluster computations were carried out using the resources of the
SKIF-Cyberia supercomputer, Tomsk State University.

\section*{References}

\end{document}